\documentclass[aoas,MSNbibl,nameyear,seceqn,dvips]{arximspdf}
\usepackage{graphicx}
\doi{10.1214/14-AOAS733} 
\volume{8}
\issue{2}
\pubyear{2014}
\firstpage{649}
\lastpage{673}

\makeatletter
\setattribute{abstract}{width}{281pt}
\newcommand{\bM}{\mathbf{M}}
\newcommand{\bK}{\mathbf{K}}
\newcommand{\bnu}{\bolds{\nu}}
\newcommand{\bZ}{\mathbf{Z}}
\newcommand{\bX}{\mathbf{X}}
\newcommand{\ba}{\mathbf{a}}
\newcommand{\bY}{\mathbf{Y}}
\newcommand{\be}{\mathbf{e}}
\newcommand{\bs}{\mathbf{s}}
\newcommand{\bphi}{\bolds{\phi}}
\newcommand{\btheta}{\bolds{\theta}}
\newcommand{\bSigma}{\bolds{\Sigma}}
\newcommand{\bepsilon}{\bolds{\varepsilon}}
\newcommand{\bdelta}{\bolds{\delta}}
\newcommand{\boldeta}{\bolds{\eta}}
\newcommand{\bxi}{\bolds{\xi}}
\makeatother

\begin{document}
\begin{frontmatter}

\title{Fast dimension-reduced climate model calibration and the effect
of data aggregation\thanksref{T1}}
\runtitle{Dimension-reduced climate model calibration}

\begin{aug}
\author[a]{\fnms{Won} \snm{Chang}\ead[label=e1]{wonchang@psu.edu}},
\author[a]{\fnms{Murali} \snm{Haran}\corref{}\ead[label=e2]{mharan@stat.psu.edu}},
\author[b]{\fnms{Roman} \snm{Olson}\ead[label=e3]{rzt2-wrk@psu.edu}}
\and
\author[b]{\fnms{Klaus} \snm{Keller}\ead[label=e4]{klaus@psu.edu}}
\runauthor{Chang, Haran, Olson and Keller}
\affiliation{Pennsylvania State University}

\address[a]{W. Chang\\
M. Haran\\
Department of Statistics\\
Penn State University\\
University Park, Pennsylvania 16802\\
USA\\
\printead{e1}\\
\phantom{E-mail:\ }\printead*{e2}}
\address[b]{R. Olson\\
K. Keller\\
Department of Geosciences\\
Penn State University\\
University Park, Pennsylvania 16802\\
USA\\
\printead{e3}\\
\phantom{E-mail:\ }\printead*{e4}}
\end{aug}
\thankstext{T1}{Supported by NSF through the Network for Sustainable Climate Risk
Management (SCRiM) under NSF cooperative agreement GEO-1240507.}

\received{\smonth{10} \syear{2013}}
\revised{\smonth{2} \syear{2014}}

%
\begin{abstract}
How will the climate system respond to anthropogenic
forcings? One approach to this question relies on climate model projections.
Current climate projections are considerably uncertain. Characterizing
and, if possible, reducing this uncertainty is an area of ongoing
research. We consider the problem of making projections
of the North Atlantic meridional overturning circulation (AMOC).
Uncertainties about climate model
parameters play a key role in uncertainties in AMOC
projections. When the observational data and the climate model
output are high-dimensional spatial data sets, the data are
typically aggregated due to computational constraints. The effects
of aggregation are unclear because statistically rigorous approaches
for model parameter inference
have been infeasible for high-resolution data. Here we develop a
flexible and computationally efficient approach using principal
components and basis expansions to study the effect of spatial data
aggregation on parametric and projection uncertainties. Our
Bayesian reduced-dimensional calibration approach allows us to study
the effect of complicated error structures and data-model
discrepancies on our ability to learn about climate model parameters
from high-dimensional data. Considering high-dimensional spatial observations
reduces the effect of deep uncertainty associated with prior
specifications for
the data-model discrepancy. Also, using the unaggregated data \mbox{results}
in sharper
projections based on our climate model. Our computationally efficient
approach may
be widely applicable to a variety of high-dimensional computer model
calibration problems.
\end{abstract}

%

\end{frontmatter}

\section{Introduction} \label{sectionintroduction}
Computer models play an important role in understanding complex physical
processes in modern science and engineering.
They are particularly important in climate science where computer
models, complex deterministic systems used to model the Earth System,
are used both to study climate phenomena as well as make projections
about the future. A major source of uncertainty in climate projections
is due to uncertainties about model parameters. Parameter calibration
involves characterizing our knowledge about a model parameter by using
observational data. Here we use calibration to refer to a statistical
method that summarizes information about a parameter in terms of a
probability distribution. In this distribution parameter values that
generate output more compatible with observational data are assigned
higher probabilities than parameters less compatible with observations.
Calibration of the parameters using observational data is hence one
avenue to reduce the uncertainty in future projections. A number of
issues and challenges arise when performing statistical calibration of
model parameters. Because each run of the computer model is
computationally expensive, computer model output is typically obtained
for a relatively small sample of parameter values. Furthermore, the
model output at each parameter setting may be high dimensional and in
the form of spatial fields. A sound statistical approach to this
problem needs to simultaneously address the spatial dependence in the
data and model outputs, account for various sources of uncertainty and
remain computationally efficient. Computational efficiency is key in
order to utilize all the relevant observations at the appropriate
scale; previous methods for climate model calibration have relied on
heavy data aggregation, thereby potentially discarding valuable information.

The scientific problem motivating our statistical analysis is the
projection of the future state of the North Atlantic meridional overturning
circulation (AMOC) in response to anthropogenic climate change.
The AMOC is a large-scale ocean circulation that transports cold and
dense water equatorward in the deep North Atlantic, and warm and salty
water poleward in the upper layers of the North Atlantic.
The AMOC might show a persistent weakening in response to
anthropogenic forcing. Because the AMOC plays an important role in
heat and carbon transport, an AMOC weakening is projected to have
considerable impacts on climate, and, in response, on natural and human
systems [cf. \citet{alley2007ipcc}, \citeauthor{keller2005avoiding} (\citeyear{keller2005avoiding,keller2007early})].
We use previously published perturbed physics ensemble runs [\citet{sriver2012toward}] of the
University of Victoria Earth System Climate Model (UVic ESCM) [\citet{weaver2001uvic}] to set up the calibration problem. Specifically, the
runs model transient behavior of the climate system over the years
1800--2100. Each run starts from a control climate, obtained by running
the model from the same initial condition to equilibrium at
preindustrial conditions.
Vertical ocean mixing is important in projecting the AMOC [\citet{wunsch2004vertical}], but most of
the mixing occurs on scales below that of the UVic ESCM,
hence, mixing is ``parameterized'' [cf. \citet{weaver2001uvic,schmittner2009using,goes2010skill}]
using a ``vertical background diffusivity'' ($K_{\mathrm{bg}}$). The AMOC
projections depend on the $K_{\mathrm{bg}}$ parameter values [e.g., \citet{goes2010skill}].
The value of $K_{\mathrm{bg}}$ is uncertain; it therefore needs to be calibrated
using observations of the climate that are informative about $K_{\mathrm{bg}}$
[cf. \citet{goes2010skill,bhatharatonkkell2009}].

Here, we calibrate $K_{\mathrm{bg}}$ using observations of ocean potential
temperature from the World Ocean Atlas 2009 [\citet{antonov2010world,locarnini2010world}]. The World Ocean Atlas is a
gridded data product generated by interpolation of instrumental
observations. The observational data at irregularly distributed
locations are interpolated onto a regular grid by Barnes interpolation
[\citet{barnes1964technique}]. The parameter $K_{\mathrm{bg}}$ affects the depth
of the oceanic
pycnocline [\citet{gnanadesikan1999simple}] and the AMOC
[\citet{bryan1987parameter,goes2010skill}]. As a consequence, models
with different $K_{\mathrm{bg}}$ values are expected to
result in different ocean temperature distributions. Ocean temperatures
are therefore informative
about $K_{\mathrm{bg}}$. Note that neither data assimilation nor calibration
play any role in producing the World Ocean Atlas data set and the data
product does not depend on any assumptions about vertical diffusivity.

Both the UVic ESCM output and the observational data are spatial data
sets with more than 60,000 spatial locations. Of particular interest is
how data aggregation affects the
calibration result for $K_{\mathrm{bg}}$. Often observations of climate and
climate model outputs are
3-D spatial fields. When the spatial data sets are large it is common
practice to aggregate them
into 1-D or 2-D patterns [\citet{Sansoforest2009,drignei2008parameter,forest2008constraining,goes2010skill,bhatharatonkkell2009,olson2012climate,schmittner2009using}]
either to avoid computational issues or
because the skill of the models at higher resolution may not always be
trusted.

An important and interesting question is what information, if any, is
lost by this data aggregation. Aggregating data for model calibration
may increase or decrease the model parameter uncertainty depending on
the relative importance of several processes. Data aggregation may lead
to information loss which would result in larger uncertainties about
the calibrated parameters. On the other hand, data aggregation can
potentially reduce the magnitude of model errors, for instance, due to
unresolved variability or structural errors in the climate models,
which could in turn lead to smaller uncertainties about the calibrated
parameters. We hypothesize for the scientific questions and models we
consider here that data aggregation may lead to considerable loss of
information resulting in increased uncertainties about model
parameters. Increased uncertainties about parameters propagates to
increased uncertainty in climate projections, which can impact risk-
and decision-analysis.

We adopt a Gaussian process-based approach to the calibration problem
[\citet{sacks1989design,kennedy2001bayesian}]. Gaussian processes
provide flexible statistical interpolators or ``emulators'' of
the computer model across various parameter settings and are therefore
attractive for climate model calibration
[cf. \citet{Sansoforest2009,bhatharatonkkell2009}]. Unfortunately, the
likelihood evaluations
involved in fitting such models can become prohibitive with
high-dimensional spatial data due to the expensive matrix operations
involved. Current approaches for high-dimensional computer model
calibration can reduce the computational burden and make likelihood
evaluation feasible for moderately large data sets (spatial fields
observed at a few thousand locations) or data sets that are on a
regular and complete grid [\citet{Higdon2008,bhatharatonkkell2009,bayarri2007computer}].
However, to our knowledge, no
current calibration approach can overcome the computational challenge
of dealing with large spatial data sets (more than tens of thousands of
data points) on an incomplete grid. Here an incomplete grid refers to
one with a large number of missing points (about~40\%).

The impact of data aggregation on climate model calibration is a
largely unanswered question due to the inability of existing methods to
analyze large spatial data sets of both computer model output and
observations. Throughout this manuscript we will use ``large'' to refer
to data sets that comprise over tens of thousands of spatial
observations. Here we develop a computationally efficient approach that
handles large data sets. This approach gives us the freedom to carry
out a careful study of the effects of data aggregation, for example,
comparing calibration based on unaggregated three-dimensional data with
calibration based on aggregated two-dimensional or one-dimensional
data. Our approach also enables one to investigate the interaction
between data aggregation and data-model discrepancies and errors when
inferring computer model parameters.
In our simulated examples, we have shown that the method can handle
complicated model-observation discrepancy processes without sacrificing
computational efficiency.


The remainder of this paper is organized as follows. In Section~\ref
{sectiondatadesc} we provide a description of the data set.
In Section~\ref{sectionframework} we describe our two-stage framework
for climate model calibration and
the associated computational challenges. In
Section~\ref{sectioncalibrationhighdim} we propose a general model
calibration approach in a reduced-dimensional space
that uses a combination of principal components and a basis
representation to overcome computational challenges.
In Section~\ref{sectionexample} we provide implementation details and
in Section~\ref{sectionresults} we discuss the results from simulated
examples and real data. We
conclude this paper with caveats and future directions in Section~\ref
{sectiondiscussion}.

\section{Data description} \label{sectiondatadesc}

Our goal is to build an emulator based on spatial output from UVic ESCM
and to calibrate vertical ocean
diffusivity ($K_{\mathrm{bg}}$) using ocean potential temperature data.
The UVic ESCM runs are 3-dimensional patterns of the
mean ocean potential temperature over 1955--2006 at 250 parameter
settings. The parameters controlling model outputs are vertical ocean
diffusivity ($K_{\mathrm{bg}}$), anthropogenic aerosol scaling factor
($A_{\mathrm{scl}}$) and
climate sensitivity ($C_{\mathrm{s}}$). Note that we converted longwave radiation
feedback factor, which is one of the original input parameters for
UVic, into $C_{\mathrm{s}}$ using a simple spline fit. We refer to \citet{sriver2012toward}
for the design points and details of the ensemble runs.

To avoid problems related to model artifacts and sparse sampling,
we excluded data beyond $60^\circ$N and $80^\circ$S and 3000~m in
depth [\citet{key2004global,schmittner2009using,bhatharatonkkell2009}].
UVic ESCM outputs are on a 77 (latitude)${}\times{}$100 (longitude)${}\times{}$13 (depth) grid, but the
number of locations that have nonmissing observations is 65,595. The
missing values occur because there is no ocean at the
locations in the UVic ESCM representation. At each grid point we
compute a temporal mean over the time period of 1955--2006 to average
out the effect of unresolved internal variability.

\begin{figure}

\includegraphics{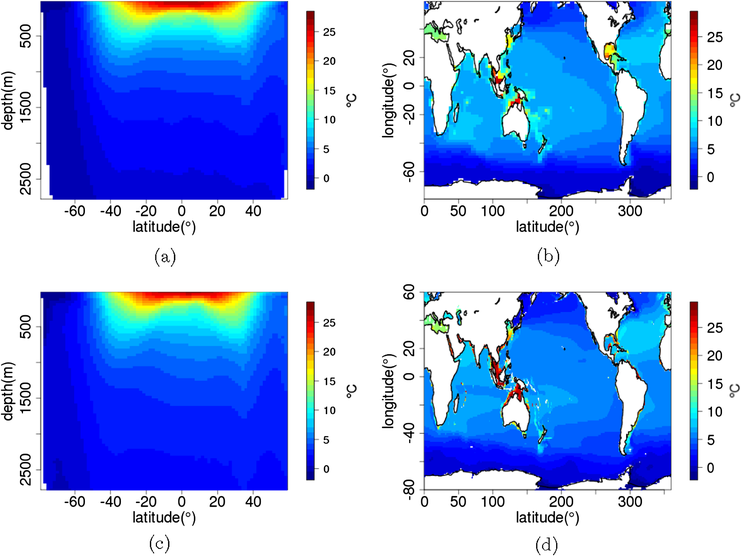}

\caption{Plots of ocean temperature patterns averaged over 1955--2006.
A UVic ESCM run (the first row)
and
the observational data from World Ocean Atlas 2009 (the second row). The left column shows the
latitude--depth profiles of zonal mean, and the right column displays
the longitude--latitude profiles of vertical mean. Note that the model
run shown here is an example of 250 model runs described in
Section~\protect\ref{sectiondatadesc}.
\textup{(a)} and \textup{(b)} Latitude--depth pattern ($K_{\mathrm{bg}}=0.2$, $A_{\mathrm{scl}}=1.5$, $C_{\mathrm{s}}=3.976$).
\textup{(c)}~and~\textup{(d)} Latitude--depth pattern (observational data).}\label{figdataex}
\end{figure}

The observational data are on a 180 (latitude)${}\times{}$360
(longitude)${}\times{}$33 (depth) grid, and we remap this observed data
into the UVic model grid using a linear interpolation using only the
nearby points. See Figures A1~and~A2 in the supplementary material [\citet{chang2013fastsupplement}] for comparison between the UVic model grid
and the observational data grid.
This results in a relatively small reduction to~61,051
data points. The model output locations are also adjusted accordingly.
We convert the observed
in situ temperature field into the potential temperature field in order
to (i) have the same measurement unit with UVic ESCM output and (ii)
adjust the effect of pressure on ocean temperature.
We obtain potential temperature from the
in situ temperature [\citet{locarnini2010world}] and salinity fields
[\citet{antonov2010world}] using the UNESCO equation of state [\citet{UNESCO1981}]
following \citet{bryden1973new} and
\citet{fofonoff1977computation}.
During the conversion procedure, we assume a simplified ocean pressure
field varying as a function of
latitude and depth [\citet{lovett1978merged}]. As we do for the UVic
ESCM output, we compute a temporal mean over 1955--2006 at each
location. Figure~\ref{figdataex} shows examples of UVic ESCM model
runs and the converted observational data.

\section{Model calibration framework} \label{sectionframework}
Our computer model calibration framework consists of two stages, (i) model
emulation and (ii) parameter calibration [\citet{bayarri2007computer,bhatharatonkkell2009}].
First, we construct an emulator that
interpolates computer model outputs at different parameter
settings using Gaussian random fields [\citet{sacks1989design}]. This
can be viewed as statistical interpolation or ``kriging'' [\citet{cressie1994}] in the computer model parameter space.
Second, we infer the computer model
parameters by relating observational data to computer model output
using the emulator, while considering observational error and
allowing for systematic discrepancies between the model and
observations [\citet{kennedy2001bayesian}]. Note that this two-stage
approach has some advantages
over fully Bayesian methods that combine the two stages into a single
inferential step. By constructing an emulator solely based on computer
model output [\citet{rougier2008comment,bhatharatonkkell2009,liu2009modularization}], this
two-stage approach ensures that
inference in the emulation stage is not contaminated by model
discrepancy and
observational error. In addition, separating the emulation stage from
calibration provides an easier way to diagnose the accuracy of an
emulator. Furthermore, computations are faster and parameter
identifiability problems are reduced.

Let $Y(\bs,\btheta)$ denote the computer model output at the spatial location
$\bs={}$(longitude, latitude, depth)${}\in\mathcal{S} \subseteq\mathbb{R}^3$
and the model parameter setting $\btheta\in\bolds{\Theta}$, where
$\mathcal{S}$ is the spatial domain of
the process and $\bolds{\Theta}$ is the computer model parameter
space, typically a subset of a unidimensional or multidimensional
Euclidean space. In our calibration problem, $\bolds{\Theta} \subset
\mathbb{R}^3$ since there are three input parameters. Furthermore,
$Z(\bs)$ is the corresponding observation at the spatial location $\bs
$. Since each run of the climate model is computationally expensive, we can
obtain computer model outputs only for a relatively small number of
design points $p$.
We denote these design points in the parameter space by
$\btheta_1,\ldots,\btheta_p \in\bolds{\Theta}$.
Let $\bY_i \in\mathbb{R}^n$ be the computer model output at each
parameter setting
$\btheta_i$ for $i=1,\ldots,p$.
Each computer model output
$\bY_i=(Y(\bs_1,\btheta_i),\ldots,Y(\bs_n,\btheta_i))^T$ is a
spatial process observed at $n$ different spatial locations $(\bs
_1,\ldots,\bs_n)$. In our calibration problem, $n={}$61,501 and $p=250$.
Let $\bY$ be the vector of concatenated computer model outputs such
that $\bY=(\bY_1,\ldots,\bY_p)^T$.
We denote the observed spatial process at $n$ locations by
$\bZ=(Z(\bs_1),\ldots,Z(\bs_n))^T$. Note that we assume that the
locations for
each model output and observation data are the same. If they are
different, one can interpolate either of them depending
on which one has a higher resolution.
Our objective is to infer the parameter
$\btheta$ by combining information from $\bZ$ and $\bY$.

\subsection{Two-stage emulation and calibration} \label{subsubsectionGRF}
We first outline our general framework for emulation and calibration.

\subsubsection*{Model emulation using Gaussian random fields}
As described in \citet{bhatharatonkkell2009}, a standard approach to
approximate the climate model output is using
a Gaussian process such that
\[
\bY\sim N \bigl(\mathbf{X}\bolds{\beta},\bSigma(\bxi_y) \bigr),
\]
with a $np \times b$ covariate matrix
$\bX$ and a vector of regression coefficient $\bolds{\beta}$.
The covariates in $\bX$ are the spatial locations (e.g., latitude,
longitude and depth) and the climate parameters. The covariate matrix
$\bX$ contains all the spatial coordinates and the parameter settings
used to define the covariance matrix $\bSigma(\bxi_y)$. The vector
$\bxi_y$ contains all the parameters determining the covariance matrix
$\bSigma(\bxi_y)$.
In our application, since the mean term $\mathbf{X}\bolds{\beta}$ is
set to $\mathbf{0}$ (see Section~\ref{sectionemulation} for more
details), we fit a Gaussian random field to $\bY$ by finding the
maximum likelihood estimate (MLE) of only $\bxi_y$, denoted by $\hat
{\bxi}_y$.\looseness=-1

The fitted Gaussian random field defines the probability model for the
computer model output at any location $\bs\in\mathcal{S}$ and
parameter setting $\btheta\in\bolds{\Theta}$. Therefore, the
Gaussian process model provides a predictive distribution of computer
model output at any untried value of $\btheta$ given the existing
output $\bY$ [\citet{sacks1989design}]. We denote the resulting
interpolated process by
$\boldeta(\btheta,\bY)$ and call it an emulator process. This
approach automatically provides a quantification of interpolation uncertainty.

\subsubsection*{Model calibration using Gaussian random field model}
Once an emulator $\boldeta(\btheta,\bY)$ is available, we model the
observational data $\bZ$,
%
\begin{equation}
\label{eqnZmulti} \bZ=\boldeta(\btheta,\bY)+\bdelta+ \bepsilon,
\end{equation}
where $\bepsilon\sim N(\mathbf{0},\sigma^2 \mathbf{I})$ is an
independently and identically
distributed observational error and $\bdelta$ is a data-model
discrepancy term. The discrepancy $\bdelta$ is also modeled as a Gaussian
process, thus, $\bdelta\sim N(\mathbf{0},\bSigma_d(\bxi_d))$ with a
spatial covariance matrix $\bSigma_d(\bxi_d)$ between the locations
$\bs_1,\ldots,\bs_n$ and
a vector of covariance parameters~$\bxi_d$. The details regarding the
specification of the covariance function are provided in Section~\ref
{subsectioncalibration}. This discrepancy term is
crucial for parameter calibration [cf. \citet{bayarri2007computer,bhat2010computer}]. Note that this problem is ill
posed without any prior
information for $\bxi_d$, so an informative prior is necessary.
Our inference for $\btheta$, $\bxi_d$ and $\sigma^2$ is based on
their resulting
posterior distribution.

\subsection{Challenges with high-dimensional data} \label{sectionchallenges}
High-dimensional data sets pose considerable computational challenges
due to the expensive likelihood function calculations that involve
high-dimensional matrix computations. For instance, in the calibration
problem described in Section~\ref{sectionexample}, the dimensionality
of the model output and the observational data is $n=984$ in the 2-D
case and $n={}$61,051 in the 3-D case, with $p=250$. The latter example involves
prohibitive computations with na\"{\i}ve implementations
(discussed
and explained in Section~A1 in the supplementary material [\citet{chang2013fastsupplement}]). For instance, with $n$-dimensional climate
model outputs at $p$ different parameter settings, evaluation of the
likelihood function requires $\mathcal{O}(n^3p^3)$ operations.
Therefore, numerical methods such as Newton--Raphson or MCMC algorithms
become infeasible.

\section{Model calibration with high-dimensional spatial data}\label{sectioncalibrationhighdim}

We develop a dimension reduction approach based on spatial basis
functions to increase computational efficiency. Spatial
basis functions can map high-dimensional data into a
\mbox{low-}dimensional space [\citet{bayarri2007computer}] and
find a representation of the probability model that results in lower
computational cost for likelihood evaluations [\citet{Higdon2008,bhatharatonkkell2009}]. Since there may be
a trade-off between parsimony and accurate inference, it is crucial to
find a set of spatial basis functions that gives a computationally
feasible likelihood formulation without considerable loss of information.
Below, we review drawbacks to the current
approaches in the context of high-dimensional spatial data and propose
a new
approach to overcome these limitations.

\subsection{Current approaches}
Various methods have been introduced to overcome computational
challenges with models for high-dimensional spatial data. These methods
may be roughly grouped into the following three categories: low-rank
representations, likelihood approximations and sparse covariance
approximations. Low-rank representation methods such as kernel
convolution [\citet{higdon1998process}], Gaussian predictive process
[\citet{banerjee2008gaussian}] and fixed rank kriging [\citet{cressie2008fixed}] approximate spatial processes using a set of basis
functions and typically reduce the computational costs by using
algorithms for \mbox{patterned} covariance matrices, for instance, the
Sherman--Morrison--Woodbury formula. Likelihood approximation methods
substitute the expensive likelihood function with a relatively
inexpensive approximation such as the Whittle likelihood [\citet{fuentes2007approximate}] or composite likelihood [\citet{vecchia1988estimation,stein2004approximating,caragea2006approximate,eidsvik2011estimation}].
Sparse covariance approximation methods such as covariance tapering
[\citet{furrer2006covariance}], Gaussian Markov random field
approximations [\citet{lindgren2010explicit,simpson2010order}] and
lattice kriging [\citet{nychka2013multi}] introduce sparsity in the
covariance or precision matrix, thereby allowing for fast computations
using sparse matrix algorithms.

A few different approaches to climate model calibration with
multivariate computer model outputs have been developed in recent
years, including \citet{bayarri2007computer,Higdon2008} and
\citet{bhatharatonkkell2009}. These approaches, however, are not
readily applicable to the 3-D model output and observations we consider
here due to the following reasons. First, in spite of the gains in
computational efficiency, likelihood evaluations remain computationally
prohibitive. The computational cost of a single
likelihood evaluation in the emulation step in
\citet{bhatharatonkkell2009} scales as $\mathcal{O} (nJ^2)$ where
$J$ is the number of knots for the kernel basis. In \citet{Higdon2008},
the computational cost scales with $\mathcal{O}(p^3J_y^3)$ where $J_y$
is the number of principal components used to represent the data.
For the 3-D calibration problem we consider, $n$ is 61,051 and $J$
should be more than
$3(K_v) \times3 (A_{\mathrm{scl}}) \times3 (C_{\mathrm{s}}) \times 3$ (depth)${}\times20$
(longitude)${} \times15$ (latitude)${}={}$24,300 to ensure the number of knots
to be at least three and greater than 20\% of the design points for
each dimension. The number of principal components $J_y$ needs be at
least 20 to have more than 90\% of explained variance. Second, the
transformation based on
the basis matrix may not be applicable to two- or three-dimensional
spatial data. Using a wavelet transformation
[\citet{bayarri2007computer}] requires the same dyadic (a power of 2)
number of data points for each spatial dimension, and the data need to
be on a regular grid without missing values; irregular data and
missing values are common in climate model calibration problems. In
addition, conducting Bayesian inference on the joint posterior
distribution may pose difficulties, both computationally as well as in
terms of prior specification and identifiability issues. For example,
\citet{Higdon2008} requires estimating $4\times J_y+1$ parameters, which
translates to an 81-dimensional distribution for the 3-D case in
Section~\ref{sectionexample}.

\subsection{Reduced-dimensional model calibration}\label{sectionourmethod}
Our method to overcome the aforementioned challenges relies on (i)
representing the spatial field using a principal component basis and
(ii) emulating each principal component separately. Instead of using a
principal component basis to reduce the complexity of matrix
computation as in
\citet{Higdon2008}, we use it to map the computer model outputs into a
low-dimensional space and construct an emulator in that space
directly. Since the principal components are uncorrelated by
construction, we can build the emulator by constructing a
1-dimensional Gaussian process for each principal component in
parallel.
Fitting Gaussian random fields for each principal component
requires estimation of only five parameters (see below). The likelihood
evaluations
involve covariance matrices of size $p\times p$. These features allow
us to construct the emulator in a computationally efficient and highly
automated manner. Moreover, since the principal component
transformation can be applied to nondyadic spatial data with
irregular locations, it has a broader range of application than
wavelet transformations. In the calibration step, we develop an
approach to map the observational data into a low-dimensional space.

\subsubsection{Computer model emulation}\label{sectionemulation}
The first step of this approach is to find the basis matrix for
computer model output.
We consider the computer model outputs as an
$n$-dimensional data set with $p$ replications and find the principal
component basis. Let $\bM$ denote the $p \times n$ matrix storing
the computer model outputs $\bY_1,\ldots,\bY_p$ such that
%
\begin{equation}
\label{eqnM} \bM=\pmatrix{ \bY_1^T
\cr
\vdots
\vspace*{2pt}\cr
\bY_p^T}.
\end{equation}
Following the standard process of finding principal components,
we first preprocess the computer model outputs to make the column
means of the matrix $\bM$ all~0's.
Applying singular value decomposition (SVD), we
find the scaled eigenvectors $\mathbf{k}_1=\sqrt{\lambda_1}
\mathbf{e}_1,\ldots, \mathbf{k}_p=\sqrt{\lambda_p} \mathbf{e}_p$, where
$\lambda_1>\lambda_2>\cdots>\lambda_p$ are the ordered eigenvalues and
$\mathbf{e}_1,\ldots,\mathbf{e}_p$ are the corresponding eigenvectors
of the covariance matrix of
$\bM$,
where $J_y \ll p$ is the number of principal components that we
decide to use in the emulator. One can choose the number of
principal components by looking at the proportion of explained
variation given by $\frac{\sum_{i=1}^{J_y} \lambda_i} {\sum_{i=1}^{p}
\lambda_i}$. We define\vspace*{1pt} the basis matrix for computer model output by
$\bK_y=(\mathbf{k}_1,\ldots,\mathbf{k}_{J_y})$.

For each parameter setting $\btheta_i$ $(i=1,\ldots,p)$, the
first $J_y$ principal components
$\bY_i^R=(Y^R_{i1},\ldots,Y^R_{iJ_{y}})^T$ are computed as
\[
\bY_i^R= \bigl(\bK_y^T
\bK_y \bigr)^{-1} \bK_y^T
\bY_i.
\]
Let $\bY^R=(\bY_1^R,\ldots,\bY_p^R)^T$, hence, each element of this matrix
$\{\bY^R\}_{ij}=Y^R_{ij}$ is the $j$th principal component at the
$i$th computer model parameter setting. Since the columns in $\bK_y$
are orthogonal, the
principal components found here are uncorrelated to
each other and this leads us to a parallelized emulator
construction that is similar to the wavelet transformation approach in
\citet{bayarri2007computer}.
For each $j$th principal component across
the parameter settings (i.e., for each $j$th column of~$\bY^R$), we
construct a Gaussian random field with the squared exponential
covariance function such that
%
\begin{equation}
\label{eqnsquaredexpcov} \operatorname{Cov} \bigl(Y^R_{kj},Y^R_{lj}
\bigr)=\kappa_{y,j}\exp \Biggl(-\sum_{i=1}^{3}
\frac{|\theta_{ik}-\theta_{il}|^2}{\phi_{y,ij}^2} \Biggr)+\zeta _{y,j} 1(\btheta_k=
\btheta_l)
\end{equation}
with partial sill $\kappa_{y,j}$, nugget $\zeta_{y,j}$ and range parameters
$\bphi_{y,j}=(\phi_{y,1j},\phi_{y,2 j},\allowbreak\phi_{y,3
j})^T$. Leave-10-percent-out
cross-validation experiments with 50 different randomly generated
parameter configurations indicate that the squared exponential
covariance shows a
better fit than alternatives such as the exponential covariance (Figure~A3 in the supplementary material [\citet{chang2013fastsupplement}]).
Note that this choice of covariance function may yield a fitted
parameter field that is too regular, resulting in imprecise estimation
of the range parameters $\phi_{y,ij}$, $i=1,\ldots,3$, $j=1,\ldots,J_y$.
However, our purpose here is not to estimate the range parameters
precisely; the range parameters depend on the distance in the parameter
space, which is a somewhat arbitrary notion.
Note also that the mean term of each Gaussian process used here is set
to be zero, since each of the principal components has zero mean across
the parameter settings.

We denote the collection of emulator parameters for the $j$th principal
component
by $\bxi_{y,j}=(\kappa_{y,j}, \zeta_{y,j}, \bphi_{y,j})^T$.
One can construct the emulator by finding
the MLE $\hat{\bxi}_{y,j}$ for each $j$ separately.
The emulator $\boldeta(\btheta,\bY^R)$ is the collection
of predictive processes of $J_y$ principal components at $\btheta$
defined by the
covariance function~(\ref{eqnsquaredexpcov}) and the MLEs
$\hat{\bxi}_{y,1},\ldots, \hat{\bxi}_{y,J_y}$.
Note that even though we construct the emulator in terms of
the principal components, we can make a projection $\bY^*$ in the original
space at a new parameter setting $\btheta^*$ by computing
\[
\bY^*=\bK_y \boldeta \bigl(\btheta^*,\bY^R \bigr).
\]
To summarize, the emulation step uses the data $\bY_1^R,\ldots,
\bY_{J_y}^R$ of dimension $p$ and computes MLEs
$\hat{\bxi}_{y,1},\ldots,\hat{\bxi}_{y,J_y}$. Hence, the
computational cost
is reduced from $\mathcal{O}(n^3p^3)$ to $\mathcal{O}(J_y p^3)$ when
compared to a na\"{\i}ve approach. The resulting fitted model is then
used for the calibration step as described in the following section.

\subsubsection{Computer model calibration}\label{subsectioncalibration}
Using $\boldeta(\btheta,\bY^R)$, the emulator for the principal
components, we
reformulate the model for observational data in (\ref{eqnZmulti}) as
\[
\bZ=\bK_y \boldeta \bigl(\btheta,\bY^R \bigr) +
\bK_d \bnu+ \bepsilon,
\]
where $\bK_d \bnu$ is a kernel convolution representation
[\citet{higdon1998process}] of the discrepancy $\bdelta$. $\bnu$ is
a\vadjust{\goodbreak}
vector of independent and identically distributed Normal random
variates at $J_d\ll
n$ locations, $\bnu\sim N(\mathbf{0},\kappa_d \mathbf{I}_{J_d})$.
$\ba_1,\ldots,\ba_{J_d} \in\mathcal{S}$. The variance parameter
$\kappa_d$ determines the magnitude of discrepancy, and the range
parameters $\phi_{d,1},\phi_{d,2} >0$ specify the bandwidth of kernels.
We define the kernel basis by
%
\begin{equation}
\label{eqnkerneldisc} (\bK_d)_{ij}=\exp \biggl(-
\frac{g(s_{1i},s_{2i},a_{1j},a_{2j})}{\phi_{d,1}} -\frac{|s_{3i}-a_{3j}|}{\phi_{d,2}} \biggr),
\end{equation}
where $s_{ki}$ and $a_{kj}$ are the $k$th elements of
$\bs_i$ and $\mathbf{a}_{j}$,
respectively.
Our choice for the knot points $\ba_1, \ldots,\ba_{J_d}$ are on a
grid of 15.6 degrees in latitude, 36 degrees in longitude and 429~m in
depth. The design of these points does not affect the resulting process
unless chosen to be too sparse.
The geodesic distance function measures the great circle
distance between two points on the Earth's surface. The function
$g(s_{1i},s_{2i},a_{1j},a_{2j})$ is given by
\[
r\arccos \bigl(\sin(s_{2i}) \sin(a_{2j}) +
\cos(s_{2i}) \cos (a_{2j}) \cos|s_{1i}-a_{1j}|
\bigr),
\]
where $r$ is the radius of Earth (6378~km).
By following \citet{Higdon2008}, the range parameters are
prespecified by scientific expert judgment; this reduces computations and
identifiability issues
(see Section~\ref{sectionexample} for specification of these
parameters). The kernel function in (\ref{eqnkerneldisc}) yields a valid
covariance under geodesic distance since it
is strictly positive definite on a sphere [\citet{gneiting2011sphere}].
We assumed separability for distance along
the surface and distance along the depth. The resulting process is
approximately twice differentiable [\citet{zhu2010estimation}], which
produces a reasonable model for discrepancy. Even though the
discrepancy model implies an isotropic discrepancy
process [\citet{higdon2002space}], the resulting process is flexible
enough to capture the general trend in the discrepancy.

Instead of using the model (\ref{eqnkerneldisc})
directly, we conduct calibration with reduced-dimensional data
for computational efficiency. Let $\bZ^{R}$ be a reduced
version of the original data such that
\[
\bZ^{R}= \bigl(\bK^T \bK \bigr)^{-1}
\bK^T \bZ=\pmatrix{ \boldeta \bigl(\btheta,\bY^R \bigr)
\cr
\bnu} + \bigl(\bK^T \bK \bigr)^{-1} \bK^T
\bepsilon,
\]
where $\bK=(\bK_y\ \ \bK_d)$. The probability model of $\bZ^R$ is
%
\begin{equation}
\label{eqnZRprob} \bZ^R\sim N \biggl( \pmatrix{\bolds{
\mu}_{\boldeta}
\cr
\mathbf{0}}, \pmatrix{ \bSigma_{\boldeta} & \mathbf{0}
\cr
\mathbf{0} & \kappa_d \mathbf{I}_{J_d}} +
\sigma^2 \bigl(\bK^T\bK \bigr)^{-1} \biggr),
\end{equation}
where $\bolds{\mu}_{\boldeta}$ and $\bSigma_{\boldeta}$ are the
mean and covariance given by the emulator $\boldeta(\btheta,\bY^R)$.
It is often helpful to apply singular value decomposition to $\bK_d$
and use the first $J_d^{PC} \ll J_d$ eigenvectors $\bK_d^{PC}$ in place
of $\bK_d$ to find $\bZ^R$. In addition to the obvious computational
advantage, this often results in better inference
since it corresponds to a regularized estimate given by ridge
regression [see \citet{trevor2009elements}, page 66]; this was
corroborated by our
extensive simulation studies.

Note that the term $\sigma^2 (\bK^T\bK)^{-1}$ in (\ref{eqnZRprob})
automatically adjusts the contribution of each principal component to
the calibration result. This can be illustrated by considering the
model without the
discrepancy, and the variance in the likelihood function is simply
$\bSigma_{\boldeta}
+\sigma^2 (\bK_y^T\bK_y)^{-1}$.
Since $ (\bK_y^T\bK_y)^{-1}$ is a diagonal matrix and its $i$th
diagonal element is the reciprocal of the $i$th eigenvalue,
$(\bK_y^T\bK_y)^{-1}$ inflates the variance of principal
components with small eigenvalues. Therefore, the principal components
with smaller explained variance will have less effect on the
calibration result.


We now briefly examine the covariance structure implied by our model.
Using the leading $J_y$ principal components, the covariance between
computer model outputs at two different spatial and parametric
coordinates $(\bs_1,\btheta_1)$ and $(\bs_2,\btheta_2)$ can be
written as
\begin{eqnarray*}\!\!
\operatorname{Cov} \bigl(Y(\bs_1,\btheta_1),Y(
\bs_2,\btheta_2) \bigr)&\approx&\operatorname{Cov} \Biggl(\sum
_{i=1}^{J_y} \sqrt{\lambda_i}
\be_{i}(\bs_1) w_i(\btheta_1),
\sum_{j=1}^{J_y} \sqrt{\lambda_j}
\be_{j}(\bs_2) w_j(\btheta_2)
\Biggr)
\\
&=& \sum_{i=1}^{J_y} \lambda_i
\be_{i}(\bs_1) \be_{i}(\bs_2)
\operatorname{Cov} \bigl(w_i(\btheta_1), w_i(
\btheta_2) \bigr),
\end{eqnarray*}
where $\be_i(\cdot)$ is the $i$th eigenfunction satisfying
\[
\int\operatorname{Cov} \bigl(Y(\theta_1,\bs_1),Y(
\theta_2,\bs_2) \bigr) \be_i(\bs
_2) \,d \bs_1 = \lambda_i \be_i
(\bs_2) \operatorname{Cov} \bigl(w_i(\btheta_1),w_i(
\btheta_2) \bigr),
\]
with the corresponding eigenvalue $\lambda_i$. 
We let $w_i(\cdot)$ denote the Gaussian process of the $i$th principal
component with the covariance function defined in (\ref
{eqnsquaredexpcov}). The leading eigenfunctions give the best
approximation among all possible orthogonal bases since it minimizes
the total mean square error [\citet{jordan1961discrete}]. 
Since we can assume different covariance functions for each principal
component process, our model can yield a nonseparable space-parameter
covariance function. In contrast, if we were to assume separability
such that
\[
\operatorname{Cov} \bigl(Y(\btheta_1,\bs_1),Y(
\btheta_2,\bs_2) \bigr)=C_{\mathrm{s}}(
\bs_1,\bs _2) C_{\btheta}(\btheta_1,
\btheta_2),
\]
for some positive definite covariance functions $C_{\mathrm{s}}$ and $C_{\theta
}$, the covariance function for the $i$th principal component process becomes
\[
\operatorname{Cov} \bigl(w_i(\btheta_1),w_i(
\btheta_2) \bigr) = C_{\theta} (\btheta _1,
\btheta_2) \lambda_i.
\]
The detailed derivation is provided in Section~A2 in the supplementary material [\citet{chang2013fastsupplement}]. The separability
assumption therefore results in a restrictive covariance structure such
that the correlation functions for all principal component processes
are the same. Hence, even though our reduced dimensional approach
utilizes a covariance that is easy to specify, it provides a richer
class of covariance functions than a separable covariance structure.
Our cross-validation studies show that our assumption is adequate for
emulating the computer model (see Section~\ref{sectionexample} for details).

\paragraph*{Priors}
We estimate the joint density of $\btheta$, $\kappa_d$
and $\sigma^2$ using the Metropolis--Hastings algorithm. Following
\citet{bayarri2007computer},
we allow for additional flexibility by estimating the partial sill parameters
$\kappa_{y,1},\ldots,\kappa_{y, J_y}$
for the emulator.
Prior specification for the parameters in the observational model
is straightforward. The discrepancy variance $\kappa_d$ and the
observational error variance $\sigma^2$ receive inverse-gamma priors
with small shape
parameter values. The prior for each parameter is a uniform
distribution over a broad range or determined by scientific knowledge.
In order to stabilize the
inference, we put an informative prior to encourage
$\kappa_{y,1},\ldots,\kappa_{y, J_y}$ to vary around their estimated values
in the emulation stage.
See Section~\ref{sectionexample} for more details about prior
specifications for our problem.

\paragraph*{Computing}
The computational costs may be summarized as follows:
\begin{longlist}[(5)]
\item[(1)] Find the basis matrix $\bK_y= ( \sqrt{\lambda_1}
\mathbf{e}_1,\ldots,\sqrt{\lambda_{J_y}} \mathbf{e}_{J_y}
)$ by computing the singular value decomposition of $\bM$ in
(\ref{eqnM}). This computation is of order
$\mathcal{O}(n^3)$, but needs to be done only once.
\item[(2)] Compute $\bY_R$ where its $i$th row is the transpose of
$(\bK_y^T\bK_y)^{-1} \bK_y^T \bY_i$.
\item[(3)] Construct a Gaussian random field for each column of $\bY_R$
by finding the MLE $\hat{\bxi}_{y,i}$ for each $i=1,\ldots,J_y$. The
computational cost is of order $\mathcal{O}(J_y p^3)$ for each
likelihood evaluation.
\item[(4)] Compute $\bZ_R=(\bK^T \bK)^{-1} \bK^T \bZ$. The computational
complexity of this step is $\mathcal{O}((J_y+J_d)^3)$.
\item[(5)] Using Metropolis--Hastings, draw an MCMC sample of $\btheta$,
$\sigma^2$, $\kappa_d$ and $\kappa_{y, 1},\ldots,\kappa_{y, J_y}$
from the
joint posterior distribution based on the model in (\ref{eqnZRprob}).
The computational cost for each likelihood evaluation is of
order $\mathcal{O}((J_y+J_d)^3)$.
\end{longlist}
The overall cost of our implementation is
$\mathcal{O}(pJ_y^3)$ for the emulation step and
$\mathcal{O}((J_y+J_d)^3)$ for the calibration step.

\section{Implementation details}\label{sectionexample}

We apply our method to data at three different aggregation levels.
In the 1-D case, we compute the
vertical means at $n=13$ different depth points [\citet{goes2010skill}]. In the 2-D case,
the zonal means are computed at $n=984$ latitude and depth
points [\citet{bhatharatonkkell2009}]. We use the original pattern
without any aggregation in the
3-D case ($n={}$61,051).
The number of principal components is determined to have more than 90\%
of the explained variation. The number of components is 5 for the 1-D, 10
for the 2-D and 20 for the 3-D case. We also tried
using 10 principal components for the 1-D, 20 for the 2-D and 30 for
the 3-D case to have more than 95\% explained variation, but did
not find any improvement in the calibration result.

\begin{figure}

\includegraphics{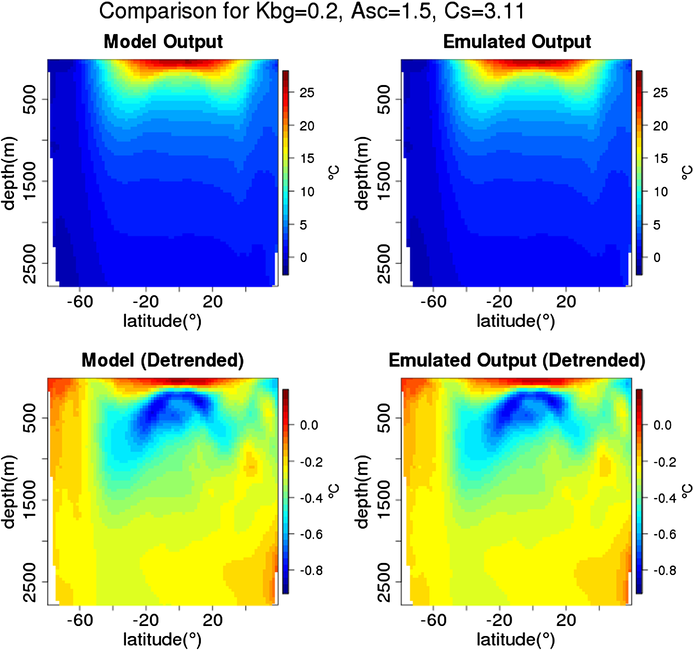}

\caption{An example of our leave-10-percent-out cross-validation
experiment for 2-D case (latitude--depth patterns). The first row shows
the comparison between the raw original output and the raw emulated
output. The second row shows the same comparison using detrended
outputs, which are computed by subtracting the mean across the
parameter settings at each location.}\label{figcrossval}
\end{figure}

We use all 250 design points in the parameter space to build the
emulator. We conducted leave-10-percent-out cross-validation and the
results show that our emulator can predict the model output precisely.
For the 2-D (latitude--depth) case, for instance, the emulator based on
principal components can reproduce well the spatial pattern at any
given parameter setting. More specifically, we randomly held out 25
model runs from the model output and then predicted these ``hold outs''
based on the remaining 225 model runs using our emulator. An example of
our results is shown in Figure~\ref{figcrossval}, which indicates that
the predicted output and the original output are essentially
indistinguishable. Other cross-validation results, including for the
3-D case, are similar. In addition, we found that the root mean squared
error was very small relative to the scale of the data. We also tested
the prediction performance of our emulator for the principal components
using uncorrelated standardized prediction errors [\citet{bastos2009diagnostics}] (see Section~A3 in the supplementary material
[\citet{chang2013fastsupplement}] for details). The graphical
diagnostics in Figure~A4 in the supplementary material [\citet{chang2013fastsupplement}] show that our emulators predict the
processes of leading principal components reasonably well.

We fix $C_{\mathrm{s}}$ and $A_{\mathrm{scl}}$ at the default values of the UVic ESCM in
the calibration stage and
make an inference only for $K_{\mathrm{bg}}$. The default values are 1 for
$A_{\mathrm{scl}}$ and 3.819
for $C_{\mathrm{s}}$. One may choose to integrate out these
two parameters, but since the ocean temperature field lacks strong
information about $A_{\mathrm{scl}}$ and $C_{\mathrm{s}}$, their estimated posterior
densities are overly dispersed. This introduces unnecessary bias in the estimate
of $K_{\mathrm{bg}}$ due to the highly nonlinear relationship between climate
parameters [\citet{olson2012climate}], thus, we decided not to
integrate out those two parameters.

Following \citet{bhatharatonkkell2009}, we assume a flat prior with a broad
range for $K_{\mathrm{bg}}$, from 0.05 to 0.55. The variance for the
observational error ($\sigma^2$) and the model discrepancy ($\kappa
_d$) receive inverse-Gamma
priors, and we denote them by $\operatorname{IG}(a_\nu, b_\nu)$ and
$\operatorname{IG}(a_z, b_z)$. We set the shape parameters for them to be
$a_\nu=2$ and $a_z=2$. To check the sensitivity of our approach to prior
specifications, we tried four different combinations, $(2,2)$,
$(2,100)$, $(100,2)$ and $(100,100)$ for $b_\nu$ and $b_z$.
The emulator variances $\kappa_{y,1},\ldots,\kappa_{y,J_y}$ also
receive inverse-Gamma priors with a shape parameter of 5. The scale
parameters are determined to have modes at their estimated values
in the emulation stage. Because of parameter identifiability problems
which in turn affected computation, we fixed the range parameter for
depth at
3000~m\vadjust{\goodbreak} and for surface as 4800~km. The knot points ($\ba_1,\ldots,\ba
_{J_d}$) are on a grid of 15.6 degrees in latitude, 36 degrees in
longitude and 429~m in depth. The design of these points does not affect
the resulting process unless chosen to be too sparse. We found that a
wide range of the
different settings for these parameters gave the same inference result
for $K_{\mathrm{bg}}$ and, hence, our particular choices did not affect the
results. The number of knot locations for the discrepancy kernel is
800 in the 3-D case, 80 in the 2-D case and 13 in the 1-D case. The
number of principal components used for the discrepancy is 200 in the
3-D case, 20 in the 2-D case and 5 in the 1-D case. The number of
principal components was determined using standard practice---by
ensuring that at least 95\% of the variability in the data was
explained in each case. 
In order to check the robustness of our results, we tried different
numbers of principal components. For example, when we increased\vadjust{\goodbreak} the
number of principal components to 300 in the 3-D case, 30 in the 2-D
case and 8 in the 1-D case we found that we obtained virtually
identical calibration results.

Finally, we note that we ran an MCMC algorithm with 25,000 iterations
for the calibration step. We carefully checked our results by comparing
summaries (e.g., posterior density estimates) based on the first 15,000
runs with those obtained from the entire 25,000 runs and verified that
our MCMC-based estimates are reliable.

\section{Results} \label{sectionresults}

\subsection{Computational benefit}
The biggest challenge in the considered analysis is the computational
cost of evaluating the likelihood function in the 3-D case, which
requires dealing with
61,051-dimensional spatial data sets. To our knowledge, current
approaches cannot address
this problem with reasonable computational effort (discussed below). In
the emulation stage, the required number of principal components is
about 20 for the 3-D case for reasonable accuracy and this means we
still need to invert a $(p\times J_y) \times(p \times J_y) = 5000
\times 5000$ covariance matrix for the likelihood evaluation in the
emulation stage if one uses the formulation due to \citet{Higdon2008}.
Moreover, the number of parameters to be estimated is $20\times4+1
=81$ ($J_y \times{}$the number of parameters for each Gaussian process
emulator${}+{}$one nugget parameter), and this also possibly increases
computational cost significantly. The method of \citet{bhatharatonkkell2009} requires multiplication of a $J \times n={}$24,300${}
\times{}$61,501 matrix into another $n \times J = \mbox{61,501}\times\mbox{24,300}$
matrix in the
likelihood evaluation, and this makes the likelihood evaluation
computationally prohibitive.

\begin{figure}

\includegraphics{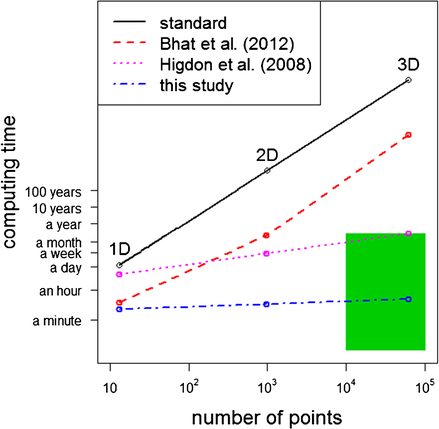}

\caption{Comparison of computational costs for the emulation step
between the current approaches and the new approach. The green box near
the bottom right corner shows computing times that are practical,
ranging from one second to three months.}\label{figcomputationalcost}
\end{figure}

The approximate computational time in emulation stage for the 1-D, 2-D
and 3-D cases for different methods are illustrated in Figure~\ref
{figcomputationalcost}. Our approximation was derived as follows:
\[
(\mbox{computing time for our approach}) \times\frac{(\mbox
{complexity for a method})}{(\mbox{complexity for our approach})}.
\]
The computing time for our approach is for the PORT routine in R [\citet{gay1983algorithm,gay1990port}] run on a system with Intel Xeon E5450
Quad-Core 3.0 GHz (without parallelization). Note that this
approximated computing time for other approaches is optimistic; we
believe that they will probably take longer than indicated. We use this
to suggest that even when viewing the other approaches' cost
optimistically, our method provides dramatically reduced computational
time. We also note here that the computing time for our approach can be
further reduced by parallelization. We describe some experimental
results on computing time reduction using parallel computing in Section
A4 in the supplementary material [\citet{chang2013fastsupplement}].

\subsection{The effect of data aggregation on climate model calibration}
\begin{figure}

\includegraphics{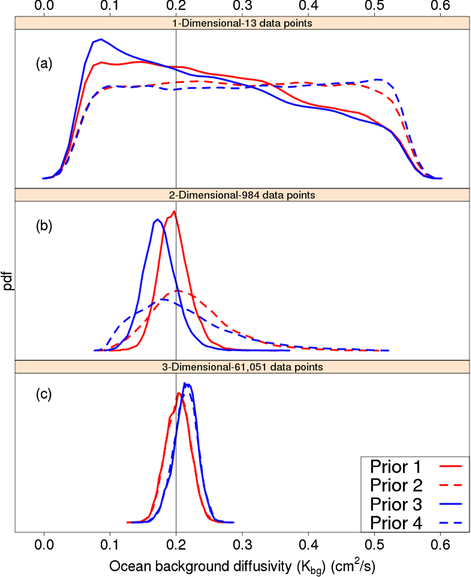}

\caption{Prior sensitivity test in the simulated example. Calibration
of $K_{\mathrm{bg}}$ value based on: \textup{(a)}~1-D depth profile,
\textup{(b)}~2-D latitude--depth pattern,
\textup{(c)}~3-D nonaggregated data. Each line represents
posterior density from four
different priors: $(b_v=2, b_z=2)$ (Prior 1, solid red line),
$(b_v=2, b_z=100)$ (Prior 2, dashed red line),
$(b_v=100, b_z=2)$ (Prior 3, solid blue line)
and $(b_v=100, b_z=100)$ (Prior 4, dashed blue line).
The solid vertical line represents the true value
of $K_{\mathrm{bg}}$ in the synthetic truth. $b_v$ and $b_z$ are
hyperparameters of $\kappa_d$ and $\sigma^2$, respectively, in (\protect\ref{eqnZRprob}).}\label{figsensitivity}
\end{figure}

\begin{figure}

\includegraphics{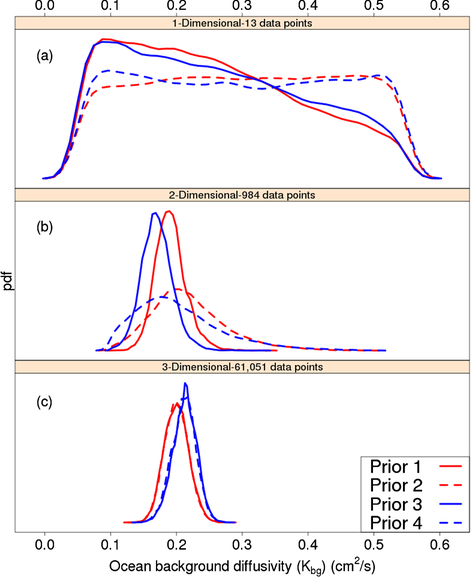}

\caption{Prior sensitivity test using observational data from the
World Ocean Atlas 2009. Calibration results based on:
\textup{(a)} 1-D depth profile,
\textup{(b)} 2-D latitude--depth pattern,
\textup{(c)} 3-D nonaggregated data. Each line
represents posterior density from four different priors:
$(b_v=2, b_z=2)$ (Prior 1, solid red line),
$(b_v=2, b_z=100)$ (Prior 2, dashed red line),
$(b_v=100, b_z=2)$ (Prior 3, solid blue line)
and $(b_v=100, b_z=100)$ (Prior 4, dashed blue line), where $b_v$ and $b_z$
are hyperparameters of $\kappa_d$ and $\sigma^2$, respectively, in
(\protect\ref{eqnZRprob}).}\label{figsensitivityreal}\vspace*{-3pt}
\end{figure}

In order to study the effect of data aggregation on climate model
calibration, we conducted a study with pseudo-observational data. The
simulated data are generated as follows:
\begin{longlist}[(3)]
\item[(1)] Choose the 3-D pattern of UVic ESCM output with $K_{\mathrm{bg}}=0.2$,
$A_{\mathrm{scl}}=1.5$ and $C_{\mathrm{s}}=3.976$ as the synthetic truth. The values for
$A_{\mathrm{scl}}$ and $C_{\mathrm{s}}$ were selected based on the fact that they were the
closest parameter values to the default values for the UVic model; the
value for $K_{\mathrm{bg}}$ was obtained from the posterior mode from previously
published work [cf. \citet{bhatharatonkkell2009}].

\item[(2)] Compute three different 3-D patterns of residuals between the
observational data (Wold Ocean Atlas 2009 data described above) and the UVic
model outputs with $K_{\mathrm{bg}}=0.1$, $K_{\mathrm{bg}}=0.2$ and $K_{\mathrm{bg}}=0.3$. The
values for $A_{\mathrm{scl}}$ and $C_{\mathrm{s}}$ are the same as in step~1. Average them
over each location to compute a pseudo-residual. This is a more
realistic and challenging residual than one obtained by simulation
from a simple error model, for example, a realization from a Gaussian
process model. For brevity, we describe here just this most
challenging case; our methods worked even better in terms of posterior
variance when error processes were assumed to be simpler.
\item[(3)] Superimpose the pseudo-residual on the synthetic truth to
construct pseudo-observational data in 3-D.
\item[(4)] Aggregate the 3-D pseudo-observational data into 2-D and
1-D by integrating the ocean temperature with respect to water volume.
\end{longlist}
The calibration results based on this simulated example are shown in
Figure~\ref{figsensitivity}. The sensitivity test indicates that the
posterior distribution of $K_{\mathrm{bg}}$ in the 1-D and the 2-D cases relies on
the specification of priors.
This deep uncertainty is drastically reduced when the full data set
(3-D) is used. A~comparison result based on the real data from Ocean
Atlas 2009 is shown in Figure~\ref{figsensitivityreal}. As in the
simulated example, the
calibration results based on the 3-D data are more robust to the prior
specification.

In addition to the experiment above, we also examined the effect of
random sampling of spatial locations, which may be considered a
reasonable alternative to\vadjust{\goodbreak} averaging over particular dimensions. As an
illustrative example we randomly chose 1300 grid points from the 3-D
locations using simple random sampling and calibrated $K_{\mathrm{bg}}$ based on
only the selected points. By repeating the same experiment with 10
different random samples, we found that calibration results are
substantially different across the samples (see Figure~A5 in the supplementary material [\citet{chang2013fastsupplement}]). These results indicate
that sampling of spatial locations may introduce additional sampling
errors to the calibration results; using all available data points is
therefore desirable.

\begin{figure}

\includegraphics{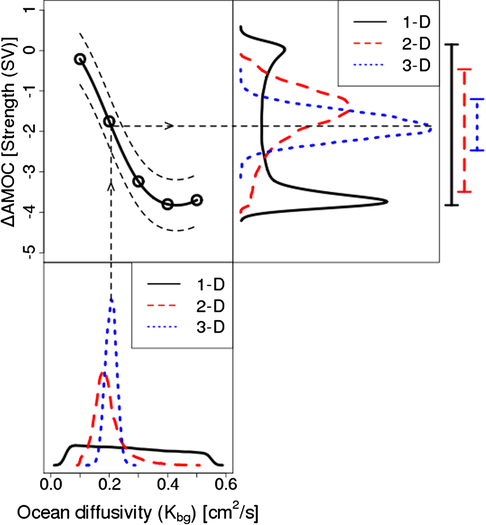}

\caption{Combined posterior densities of $K_{\mathrm{bg}}$ from different prior
specifications (lower left), the
relationship between $K_{\mathrm{bg}}$ and projected AMOC change of the 2070--2099 mean from
the 1970--1999 mean (upper left),
and the resulting AMOC change projections (upper right) using 1-D (solid black
line), 2-D (dashed red line) and 3-D (dotted blue line) data with their 95\%
credible intervals (bars at the right).}\label{figamocprojection}
\end{figure}

Using the full pattern of the 3-D data has important benefits, as it
drastically reduces the deep uncertainty due to different prior specifications.
We hypothesize that this is because the full nonaggregated spatial patterns
contain more information about both the observational error and the discrepancy.
In order to reflect the
uncertainty due to prior choice to our density estimate for $K_{\mathrm{bg}}$, we
show in Figure~\ref{figamocprojection} the posterior distributions
when the prior is assumed to be with equal probability any one of the
4 priors considered, along with the resulting AMOC projections. We
define AMOC projection as the annual maximum value of the meridional
overturning streamfunction in the Atlantic between $0^\circ$ and
$70^\circ$N. The corresponding projections for AMOC change between
1970 to 1999
mean and 2070 to 2099 mean indicate that the unaggregated pattern gives a
much narrower 95\% predictive interval than the aggregated
ones. Therefore, data aggregation increases the deep uncertainty
surrounding AMOC projection, and using unaggregated data reduces
uncertainty regarding the future behavior of the AMOC.

\section{Discussion} \label{sectiondiscussion}
\subsection{Summary}

We have considered here the problem of
calibrating a climate model parameter, $K_{\mathrm{bg}}$, in an Earth System
Model of
Intermediate\vadjust{\goodbreak} complexity by using spatial observations of the potential
temperature. In order to study the effects on both calibration and
climate projections of using unaggregated versions of the data, we
develop new methods that are computationally tractable for calibration
with high-dimensional spatial data sets. Using our methods, we show
that utilizing 3-D spatial data reduces
the uncertainty about $K_{\mathrm{bg}}$ and is more robust to various prior
specifications than calibration based on 2-D or 1-D aggregated
versions of the data. The results suggest that using unaggregated
data is valuable for reducing deep uncertainty associated with
different priors. We note that we have tested our method in several
other real data calibration problems and have obtained similar results.
For example, we carried out calibration for $K_{\mathrm{bg}}$ using a completely
different set of observations, CFC-11 [\citet{bhatharatonkkell2009}],
and found again that using the 3-D pattern sharpened our inference about
$K_{\mathrm{bg}}$ when compared to 2-D patterns.

We have demonstrated here that our computer model calibration approach is
computationally efficient even when dealing with
high-dimensional data. By exploiting the orthogonality of a principal
components decomposition of the data, this method can keep the computational
cost affordable for high-dimensional data with more than 60,000
spatial locations and 250 parameter settings. In addition, our
simulated examples show that
our approach can handle complicated model-observation
discrepancies. The method can be easily extended to allow for
calibration with multiple tracers---we can simply consider the
variance--covariance matrix for all tracers and use its principal
components to build an emulator. 
For the 2-D case, the posterior densities from using a separable
covariance structure are not very different from our PCA-based
approach; in fact, the bias appears to be slightly smaller for our
approach (Figure~A6 in the supplementary material [\citet{chang2013fastsupplement}]). Moreover, our method results in a sharper
density, which is an important criterion for calibration performance
[cf. \citet{gneiting2007probabilistic}]. For larger data, it is more
challenging to devise a simple but flexible approach that scales as
well as the PCA-based method.
We note here that our approach can be applied to even larger data sets.
In climate science, data sets consisting of millions of data points are
common. Our calibration approach in principle applies immediately to
such data. A potential computational bottleneck is the SVD. However, we
note that the SVD needs to be performed only once and we sidestep
issues related to memory management since we do not have to store the
large matrices. Furthermore, high-dimensional SVD is an active area of
research and, hence, computationally efficient approaches are being
developed [cf. \citet{halko2011finding,irlba}].

In the context of computer model calibration for making climate
projections in general, we find that spatial data aggregation appears
to generally lead to larger uncertainties (Figure~\ref{figamocprojection}), which is consistent with what we had hypothesized
about the model and observations. This implies that the effect of
information loss on the calibrated parameter is more important than the
effect of reduction in model errors by aggregation. From a climate
projections and decision-making perspective, the projections we obtain
can be used as an input to integrated assessment or economic models;
reduced uncertainties may therefore have tangible implications. Our
method is immediately available for use with other climate models and
observational products. By virtue of unlocking the full wealth of
previously untapped information in large three-dimensional data sets,
our method has a strong potential to improve projections of a host of
policy-relevant climate variables.

\subsection{Caveats}
A general issue with principal components is also worth considering in
this context: the principal components for the computer model outputs
are selected based on explained
variation and, thus, there is no guarantee that these leading principal
components carry the most important information about the climate
parameters. However, our extensive study of the effect of changing the
number of principal components suggests that this is not problematic in
our context. Our results are consistent with the recent theoretical
results in \citet{artemiou2009principal} that suggest that there is a
low probability that other (nonleading) principal components will have
a strong correlation with the climate parameters.
We hypothesize that our principal components-based approach does not
lose valuable information about the climate parameters. In the
discrepancy model, one
important simplifying assumption is the separability between surface and
depth effects. Our simulated example shows that the separability
assumption provides a good approximation to the realistic discrepancy
process. Nonseparable covariance
function that combines geodesic distance and Euclidean distance remains
as an avenue of ongoing and alive research and the subject of future
work. Furthermore, our study of calibration with simulated examples
shows that even though the number of $K_{\mathrm{bg}}$ settings at which the
model is run is relatively sparse, there is enough information to
reliably calibrate $K_{\mathrm{bg}}$ based on our emulator.

Our study is also subject to the usual caveats with respect to
scientific conclusions. First, we ignore the interpolation uncertainty
when we compute the density of AMOC projection based on the density of
$K_{\mathrm{bg}}$. Second, the result is based on a single data set and, thus,
we cannot fully evaluate the effect of structural uncertainty due to
the model-observation discrepancy and
unresolved natural variability cannot be accounted for; this
variability could impact conclusions as well
[\citet{olson2013OSSE}]. These caveats, of course, apply to almost all
existing approaches to climate model calibration and projection.

\section*{Acknowledgments}
The authors are very grateful to an anonymous referee, an Associate
Editor and the Editor for very detailed and helpful comments and
suggestions that have greatly improved this manuscript.
All
views, errors and opinions are solely that of the authors.

\subsection*{Author contributions}
Won Chang and Murali Haran formulated the statistical method. Won Chang wrote all the computer
code for emulation and calibration and wrote the first draft of the
manuscript. Murali Haran edited the text. Klaus Keller designed the AMOC projection study
and edited the text. Roman Olson supplied the previously-published UVic ESCM
model runs and processed observational data from World Ocean Atlas 2009.

\begin{supplement}
\stitle{Supplement to ``Fast dimension-reduced climate model calibration and the effect of data aggregation''}
\slink[doi]{10.1214/14-AOAS733SUPP}
\sdatatype{.pdf}
\sfilename{AOAS733\_supp.pdf}
\sdescription{Further technical details and supplementary figures can be found in this supplementary material.}
\end{supplement}


%

\printaddresses

\end{document}